\documentstyle[11pt,fleqn]{article}
\newcommand{\preprint}[1]{\hfill{\sl preprint - #1}\par\bigskip\par\rm}
\renewcommand{\title}[1]{\begin{center}\Large\bf #1\end{center}\rm\par\bigskip}
\renewcommand{\author}[1]{\begin{center}\Large #1\end{center}}
\newcommand{\address}[1]{\begin{center}\large #1 \end{center}}

\def\dinfn{\smallskip Dipartimento di Fisica, Universit\`a di Trento\\
                           and Istituto Nazionale di Fisica Nucleare,\\
                                   Gruppo Collegato di Trento, Italia}

\def\Idinfn{\address{\dinfn}}
\def\references{\begin{thebibliography}{10}}
\def\endreferences{\end{thebibliography}}
\newcommand{\s}[1]{\section{#1}\renewcommand{\theequation}
        {\mbox{\arabic{section}.\arabic{equation}}}\setcounter{equation}{0}}
\renewcommand{\ss}[1]{\subsection{#1}}

\newcommand{\app}[1]{\section{#1}\renewcommand{\theequation}
        {\mbox{\Alph{section}.\arabic{equation}}}\setcounter{equation}{0}}
\renewcommand{\date}[1]{\par\bigskip\par\sl\hfill #1\par\medskip\par\rm}
\newcommand{\email}[1]{e-mail: \sl #1@science.unitn.it\rm}
\newcommand{\femail}[1]{\footnote{\email{#1}}}
\newcommand{\pacs}[1]{\smallskip\noindent{\sl PACS number(s):
                       \hspace{0.3cm}#1}\par\bigskip\rm}
\def\babs{\hrule\par\begin{description}\item{Abstract: }\it}
\def\eabs{\par\end{description}\hrule\par\medskip\rm}
\newcommand{\ack}[1]{\par\section*{Acknowledgments} #1}
\renewcommand{\vec}[1]{{\bf #1}}       
\def\M{{\cal M}}                       
\def\hs{\qquad\qquad}         
\def\nn{\nonumber}            
\def\beq{\begin{eqnarray}}    
\def\eeq{\end{eqnarray}}      
\def\at{\left(}               
\def\aq{\left[}               
\def\ag{\left\{}              
\def\ct{\right)}              
\def\cq{\right]}              
\def\cg{\right\}}             
\def\R{\mbox{$I\!\!R$}}                 
\def\Z{\mbox{$Z\!\!\!Z$}}               
\def\ii{\infty}                         
\newcommand{\fr}[2]{\mbox{$\frac{#1}{#2}$}}      
\def\Tr{\,\mbox{Tr}\,}                  
\def\PP{\,\mbox{PP}\,}                  
\def\Res{\,\mbox{Res}\,}                
\renewcommand{\Re}{\,\mbox{Re}\,}       
\renewcommand{\Im}{\,\mbox{Im}\,}       
\def\lap{\Delta}                        
\def\al{\alpha}
\def\be{\beta}
\def\ga{\gamma}
\def\de{\delta}
\def\ep{\varepsilon}
\def\ze{\zeta}

\def\ka{\kappa}
\def\la{\lambda}
\def\ro{\varrho}
\def\si{\sigma}

\def\Ga{\Gamma}

\def\La{\Lambda}

\begin{document}

\preprint{UTF 357}
\title{
Finite Temperature Effects for Massive Fields \\
in  $D$-dimensional Rindler-like Spaces}
\author{Andrei A. Bytsenko\footnote{email: root@fmf.stu.spb.su
(subject: Prof. A.A. Bytsenko)}}
\address{State Technical University, St. Petersburg 195251, Russia}
\author{Guido Cognola\femail{cognola} and
Sergio Zerbini\femail{zerbini}}
\Idinfn

\date{August 1995}

\babs
The first quantum corrections to the free energy for  massive
fields in $D$-dimensional space-times of the form
$\R\times\R^+\times\M^{N-1}$, where $D=N+1$ and $\M^{N-1}$
is a constant curvature manifold, is investigated by means of the
$\zeta$-function regularization.
It is suggested that the nature of the divergences, which are
present in the thermodynamical quantities, might be better
understood making use of the conformal related
optical metric and associated techniques. The general form of the
horizon divercences of the free energy is obtained as a function of
free energy densities of fields having negative square masses (absence
of the gap in the Laplace operator spectrum) on ultrastatic manifolds
with hyperbolic spatial section $H^{N-2n}$ and of the Seeley-DeWitt
coefficients of the Laplace operator on the manifold $\M^{N-1}$.
Furthermore, recurrence relations are found
relating higher and lower dimensions.
The cases of Rindler space,
where $\M^{N-1}=\R^{N-1}$ and very massive $D$-dimensional black holes,
where $\M^{N-1}=S^{N-1}$ are treated as examples.
The renormalization of the internal energy is also discussed.
\eabs

\pacs{04.62.+v, 04.70.Dy}

\s{Introduction}
Recently there has been a renewed interest in the physics of black
holes. Several issues like the interpretation of the
Bekenstein-Hawking classical formula for the black hole entropy, the puzzle
of loss of information in the black hole evaporation
and the interpretation of the Hawking temperature have been discussed
(see, for example, the review \cite{beke95b}).
Furthermore in many papers, it has been pointed
out that it should be desirable
to have the usual statistical interpretation of the
black hole entropy as the number of the gravitational states at the
horizon and to try to understand the dynamical origin of the black
hole entropy (see, for example \cite{barv95-51-1741}).

On general grounds, the density of levels as a function of
the mass $M$, of a $D$-dimensional black hole should read
 \cite{harm93-47-2438} (for details, see Sec. 6.3)
\beq
\Omega(M)\simeq C_D(M)
\exp\at\frac{4\pi\hat{G}_D}{D-2}M^{\frac{D-2}{D-3}}\ct
\:.\label{td}
\eeq
where $C_D(M)$ is a quantum prefactor and $\hat G_D$ is related
to the generalized Newton constant (see Eq.~(\ref{rh})).
For a 4-dimensional black hole we have \cite{thoo85-256-727}
\beq
\Omega(M)\simeq C_4(M)\exp(4\pi G M^2)
\:,\label{t}
\eeq
$4\pi GM^2$ being the Bekenstein-Hawking classical entropy
\cite{beke73-7-2333,hawk75-43-199,gibb77-15-2752}.
As 't Hooft has pointed out, the prefactor $C_4(M)$ (the first
quantum correction to the classical result), which is usually computable for
quantum fields or extended objects (such as strings or p-branes)
in an ultrastatic space-time background, turns out to be divergent.
This prefactor may be regarded as the contribution
associated with the first quantum correction to the classical free energy.
Several different methods have been used in dealing
with such divergences, for example  "the brick wall method"
\cite{thoo85-256-727,suss94-50-2700,ghos94-73-2521}, the conical
singularity method
\cite{bana94-72-957,dowk94-11-55,solo94u-246,furs95-10-649,call94-333-55,kaba94-329-46},
critically discussed in \cite{empa95-51-5716,deal95u-33}
and the related "entaglement entropy method"
\cite{bomb86-34-373,sred93-71-666,lars94u-89}.

The horizon divergences have also been associated with the information loss
issue of black holes \cite{thoo85-256-727,suss94-50-2700} and their
physical origin, for quantum fields
\cite{isra76-57-107,barb94-50-2712} or strings
\cite{dabh95-347-222,eliz95-10-1187,empa94u-3,byts95u-130},
may be described by the following simple considerations.

In a $D$-dimensional static space-time with horizons,
the equivalence principle implies that a system in thermal
equilibrium has a local Tolman temperature  given by
$T(x)=T/{\sqrt {-g_{00}(x)}}$, $T$ being the asymptotic temperature.
Roughly speaking, near the canonical horizon
(this means that the quantity $g_{00}$ has simple zeros),
a static space-time may be regarded as a Rindler-like space-time.
We will show that, if one denotes by $\rho$ the proper
distance from the horizon, one gets for the Tolman temperature
$T(\rho)=T/\rho$. As a consequence, the total entropy for a quantum
bosonic gas reads (omitting a multiplicative constant)
\beq
S\equiv\int d{\vec x}\int_0^\ii T(\rho)^{D-1}\,d\rho \simeq
A T^{D-1}\int_0^\ii\rho^{-D+1}\,d\rho
\:,\label{mmm}
\eeq
where $A$ is the integral on the transverse coordinates $\vec x$,
namely the area of the horizon.
The latter integral is clearly divergent.
Introducing a horizon cutoff parameter $\ep$ we may rewrite it as
\beq
S&\simeq & AT\int_\ep^\La \rho^{-1}\,d\rho
\simeq AT\ln \frac{\La}{\ep} \, , \hs\mbox{for }D=2\,\, \La\,\,
\mbox{infrared cutoff}   \nn \\
S&\simeq & AT^{D-1}\int_\ep^\ii\rho^{-D+1}\,d\rho\simeq
\frac{AT^{D-1}}{(D-2)\ep^{D-2}}\, , \hs\mbox{for } D>2
\:.\label{nnn}\eeq
For the sake of generality, we write down the asymptotic high
temperature expansion for  the entropy of a quantum gas
on a $D=N+1$-dimensional static space-time defined by the metric
\beq
ds^2=g_{00}(\vec{x})(dx^0)^2+g_{ij}(\vec{x})dx^idx^j\:,
\hs \vec{x}=\{x^j\}\:,\hs i,j=1,...,N\:,
\label{x}
\eeq
$g(\vec{x})$ denoting its  determinant.
Again the equivalence principle leads to
\beq
S\simeq
T^{N}\int\at\frac{g(\vec{x})}{g_{00}(\vec{x})}
\ct^{-N/2}\,dx^{N}\,.
\label{k}
\eeq
As a consequence, the horizon divergences depend on the
nature of the poles of the integrand $(g/g_{00})^{-N/2}$.
In general, for extremal black holes, where  $g_{00}$ has higher order
zeroes, the divergences are much more severe than in the non extremal
case (see for example Refs.~\cite{mitr95u-42,deal95u-33,cogn95u-348}).

These considerations suggest the use of the optical metric
$\bar{g}_{\mu\nu}=g_{\mu\nu}/g_{00}$, conformally
related to the original one, in order to
investigate these issues. It is our
opinion that the conformal transformation techniques are particularly
suitable for studying finite temperature effects for fields in
space-times with horizons and here we would like to present some examples
of computation. This method has already been appeared for example in
Refs.~\cite{dowk78-11-895,page82-25-1499,brow85-31-2514} and has been
recently used in the horizon divergence problems in
Refs.~\cite{barb94-50-2712,barv95-51-1741,empa95-51-5716,deal95u-33,barb95u-155,bord94u-54}.
See also \cite{deal94u-347}, where the same
result is obtained with a different approach.

One of the purposes of this paper is to implement this idea in the case of
massive fields in $D$-dimensional Rindler-like space-times, we are going
to introduce.
Let us consider static space-times admitting canonical horizons and
having the topology of the form $\R\times\R^+\times\M^{N-1}$.
The metric reads
\beq
ds^2=-\frac{b^2\rho^2}{r_H^2} dx_0^2+d\rho^2+d\si^2_{N-1}\, ,
\label{rl}
\eeq
where $r_H$ is a dimensional constant, $b$ a constant factor and
$d\si^2_{N-1}$  the spatial metric related to the
$N-1$-dimensional manifold $\M^{N-1}$.
If $\M^{N-1}\equiv\R^{N-1}$, $b=1$ and  $r_H=1/a$,
$a$ being a constant acceleration, one has to deal with the Rindler space-time.
Quantum fields in such a case have been considererd in many places,
see for example
Refs.~\cite{cand76-17-2101,haag84-94-219,free85-255-693,birr82b,more95u-52}.
If $\M^{N-1}\equiv S^{N-1}$, $b=(D-3)/2$ and $r_H$ is the Schwarzschild
radius of a black hole, then we shall show that one is dealing
with a space-time which approximates, near the horizon and in the
large mass limit, a $D$-dimensional black hole (see Sec.~\ref{S:BH}).

It is well known that space-times with
canonical horizons admit a distinguished temperature, the
(Unruh) Hawking temperature. There are several ways to compute it, one
of the simplest makes use of the relation with the related surface
gravity. The other one consists in imposing the absence of conical
singularities in the Euclidean continuation of the space-time
itself \cite{gibb77-15-2752}.
For the metric (\ref{rl}), one obtains
\beq
\be_H=\frac{2 \pi r_H}{b}\,.
\label{ht}
\eeq
One can arrive at the same result working without using the Euclidean
continuation method, but making use of the principle of local
definiteness in quantum field theory \cite{haag84-94-219,more95u-50}.
Note that these method are no longer equivalent when one is dealing
with extremal black holes \cite{ghos95u-32}.
It is also important to stress that the variable $\rho$ defined by the
metric in Eq.~(\ref{rl}) has the meaning of radial proper distance
between the horizon and a point outside it and so, the divergences
of thermodynamical quantities are automatically expressed in an
invariant way.

The contents of the paper are the following.
In Sec.~\ref{S:CTT}, a review of the
necessary conformal transformation techniques is presented.
In Sec.~\ref{S:SF}, we consider a Laplace-type
operator defined on a class of $D=N+1$-dimensional space-times,
whose spatial sections have metrics
conformally related to $\M^N=\R^+\times\M^{N-1}$.
Since in general $\M^N$ in a non compact manifold,
the Laplace operator has a continuum spectrum and a general form
for the Plancherel measure, which is the analogue of the
degeneracy in the case of discrete spectrum, is presented.
The measure is used in Sec.~\ref{S:HK} in order to obtain a useful
form for the trace of the heat kernel, which is necessary for the
derivation of the free energy, which we derive in Sec.~\ref{S:FE}.
It is pointed out that in the Rindler case,
the spatial section of the conformally related space-time
turns out to be an $N$-dimensional hyperbolic manifold. In this case,
the massless scalar field can be treated without approximations.
In Sec.~\ref{S:PA} some applications
to the statistical mechanics of a scalar field
in $D$-dimensional Rindler and black hole space-times are presented
and the divergences of the first quantum corrections
to free energy and entropy are given.
Finally, we end with some conclusions in Sec.~\ref{S:C}
and with a resume of heat kernel, $\zeta$-function and
free energy on constant curvature manifolds in the Appendix.

\s{Conformal transformation techniques and optical manifold}
\label{S:CTT}

In this section we shall briefly summarize the method of
conformal transformations using $\zeta$-function regularization
\cite{dowk78-11-895,gusy87-46-1097,dowk88-38-3327,dowk89-327-267}.
These techniques permit to compute all physical quantities in an ultrastatic
manifold (called the optical manifold \cite{gibb78-358-467}) and,
at the end of calculations, transform  back them to a static one.
This method is particularly useful in
dealing with finite temperature effects for quantum fields, since these
effects can be easily investigated in ultrastatic space-times.

To start with, we consider a non self-interacting scalar field
on a $D=N+1$-dimensional static space-time defined by the metric
(\ref{k}), i.e.
\beq
ds^2=g_{00}(\vec{x})(dx^0)^2+g_{ij}(\vec{x})dx^idx^j\:,
\hs \vec{x}=\{x^j\}\:,\hs i,j=1,...,N\:.
\eeq

The one-loop partition function is given by (we perform the Wick
rotation $x_0=-i\tau$, thus all the differential operator one is
dealing with will be elliptic)
\begin{equation}
Z=\int d[\phi]\,
\exp\at-\frac12\int\phi L_D \phi d^Dx\ct
\:,\end{equation}
where $\phi$ is a scalar density of wight $-1/2$ and
the operator $L_D$ has the form
\beq
L_D=-\lap_D^g+m^2+\xi R^g
\:.\eeq
Here $m$ (the mass) and $\xi$ are arbitrary parameters, while
$\lap_D^g$ and $R^g$ are respectively the Laplace-Beltrami operator
and the scalar curvature of the manifold in the original metric $g$.

The ultrastatic metric $\bar{g}_{\mu\nu}$ can be related to the static one by
the conformal transformation
\beq
\bar{g}_{\mu\nu}(\vec{x})=e^{2\si(\vec{x})}g_{\mu\nu}(\vec{x})
\:,\eeq
with
$\si(\vec{x})=-\frac{1}{2}\ln g_{00}$. In this manner,
$\bar{g}_{00}=1$ and $\bar{g}_{ij}=g_{ij}/g_{00}$ (optical metric).
Recalling that by a conformal transformation (we remind that $\phi$ is
a scalar density)
\beq
R^{\bar g}&=&e^{-2\si}\aq R^g-2(D-1)\lap_D^g\si
-(D-1)(D-2)g^{\mu\nu}
\partial_{\mu}\si\partial_{\nu}\si\cq\:,\nn\\
\bar\phi&=&e^{\si}\phi\:,\\
\lap_D^{\bar g}\bar\phi&=&e^{-\si}\aq\lap_D^g
-\frac{D-2}{2}\lap_D^g\si
-\frac{(D-2)^2}{4}g^{\mu\nu}
\partial_{\mu}\si\partial_{\nu}\si
\cq\phi\nn\\
&=&e^{-\si}\aq\lap_D^g+\xi_D(e^{2\si}R^{\bar g}-R^g)
\cq\phi\nn
\:,\label{CT}\eeq
one obtains
\beq
L_D\phi=e^{\si}\ag-\lap_D^{\bar g}+\xi_DR^{\bar g}
+e^{-2\si}[m^2+(\xi-\xi_D)R^g]\cg\bar\phi
\:,\eeq
where $\xi_D=(D-2)/4(D-1)$ is the conformal invariant factor.
{}From the latter equation we have
$\phi L_D\phi=\bar\phi\bar L_D\bar\phi$, where, by definition
\beq
\bar L_D=e^{-\si}L_De^{-\si}=-\lap_D^{\bar g}+\xi_DR^{\bar g}
+e^{-2\si}\aq m^2+(\xi-\xi_D)R^g\cq
\label{Aq}
\:.\eeq
This means that action $\bar I=I$ by definition.
Note that classical conformal invariance requires the action
to be invariant in form, that is $\bar I=I[\bar\phi,\bar g]$,
as to say $\bar L_D=L_D$. As is well known,
this happens only for conformally coupled massless fields ($\xi=\xi_D$).

For the one-loop partition function we have
\beq
\bar Z=J[g,\bar g]\,Z
\:,\eeq
where $J[g,\bar g]$ is the Jacobian of the conformal transformation.
Such a Jacobian can be computed for any infinitesimal conformal
transformation  \cite{gusy87-46-1097}. To this aim it is
convenient to introduce a family of continuous conformal transformations
\beq
g^q_{\mu\nu}=e^{2q\si}g_{\mu\nu}
=e^{2(q-1)\si}\bar g_{\mu\nu}\:,\hs
\sqrt{g^q}\equiv\sqrt{|\det g^q_{\mu\nu}|}=e^{Dq\si}\sqrt{g}
\:,\eeq
in such a way that the metric is $g_{\mu\nu}$ or $\bar{g}_{\mu\nu}$
according to whether $q=0$ or $q=1$ respectively.
Then one gets
\beq
\ln J[g_q,g_{q+\de q}]
=\ln\frac{Z_{q+\de q}}{Z_q}
=\frac{\de q}{(4\pi)^{D/2}}
\int k_D(x|L_D^q)\si(x)\sqrt{g^q}d^Dx
\label{deJ}
\:,\eeq
where $k_D(x|L_D^q)$, is the Seeley-DeWitt coefficient, which in the case of
conformal invariant theories, is proportional to the trace anomaly.
In general, one has the asymptotic expansion
\beq
\Tr e^{-tL_D}\simeq\sum_n K_n(L_D)t^{\frac{n-D}{2}}\:,\hs
K_n(L_D)=\frac1{(4\pi)^{D/2}}\int_{\M} k_n(x|L_D)\sqrt{g}\,d^Dx
\:.\label{hk}
\eeq
If the manifold is without boundary then $K_n=0$ for any odd $n$.
The heat kernel coefficients are computable and depend on invariant quantities
built up with curvature (field strength) and their derivatives (see,
for example, \cite{bran90-15-245}).

The Jacobian for a finite transformation can be obtained from
Eq.~(\ref{deJ}) by an elementary integration in $q$
\cite{gusy87-46-1097}. In particular we have
\beq
\ln J[g,\bar{g}]=\frac{1}{(4\pi)^{D/2}}
\int_0^1dq\int k_D(x|L_D^q)\si(x)\sqrt{g^q}\,d^Dx
\:,\label{lnJ}\eeq
and finally,  making use of the $\zeta$-function
regularization, one has
\beq
\ln Z=\ln\bar Z-\ln J[g,\bar g]
=\frac{1}{2}\ze'(0|\bar L_D\ell^2)-\ln J[g,\bar{g}]
\:,\label{lnZ-Zbar}\eeq
where $\ell$ is an arbitrary parameter necessary to adjust the
dimensions and $\ze'$ represents the derivative
with respect to $s$ of the $\zeta$-function
$\ze(s|\bar L_D\ell^2)$ related to the operator
$\bar L_D$, which is given by Eq.~(\ref{Aq}).

The same analysis can be easily extended to the finite temperature
case \cite{dowk88-38-3327}. In fact we recall that for a scalar field  in
thermal equilibrium at
finite temperature $T=1/\be$ in an ultrastatic space-time, the corresponding
partition function $\bar{Z}_\be$
may be obtained, within the path integral approach, simply by Wick rotation
$\tau=ix^0$ and imposing a $\be$ periodicity in $\tau$ for
the field $\bar\phi(\tau,x^i)$ ($i=1,...,N$, $N=D-1$)
\cite{bern74-9-3312,dola74-9-3320,wein74-9-3357,kapu89b}.
In this way, in the one loop approximation, one has
\begin{equation}
\bar{Z}_\be=\int_{\bar\phi(\tau,x^i)=\bar\phi(\tau+\be,x^i)}
d[\bar\phi]\,\exp\at-\int_0^\be
d\tau\int\bar\phi\bar L_D\bar\phi\,d^Nx\ct
\:.\label{PF}
\end{equation}
in which
\beq
\bar L_D=-\partial_\tau^2-\bar\lap_N+\xi_DR^{\bar g}
+e^{-2\si}\aq m^2+(\xi-\xi_D)R^g\cq
=-\partial_\tau^2+\bar{L}_N
\:.\label{aconf}\eeq

Since the space-time is ultrastatic, by means of the $\zeta$-function
regularization again one easily obtain \cite{byts94u-325}
\beq
\ln\bar Z_\be&=&-\frac{\be}{2}\aq
\PP\ze(-\fr12|\bar{L}_N)
+(2-2\ln2\ell)\Res\ze(-\fr12|\bar{L}_N)\cq \nn \\
&&\hs+\lim_{s \to 0}\frac{d}{ds} \frac{\be}{\sqrt{4\pi}\Ga(s)}
\sum_{n=1}^\ii\int_0^\ii t^{s-3/2}\,e^{-n^2\be^2/4t}\,
\Tr e^{-t\bar{L}_N}\,dt
\label{logPF-Jacobi}\:.
\eeq
where $\PP$ and $\Res$ stand for the principal part and the residue
of the function.
The free energy is related to the canonical partition function
by means of equation
\beq
F(\be)=-\frac{1}{\be}\ln Z_\be
=-\frac{1}{\be}\at\ln\bar Z_\be-\ln J[g,\bar g]\ct
=F_0+F_\be\:,\label{FE}\eeq
where $F_0$ represents the vacuum energy,
which is given by the first term in
Eq.~(\ref{logPF-Jacobi}), while $F_\be$ represents the
temperature dependent part (statistical sum). The entropy and the
internal energy of the
system are given by the usual thermodynamical formulae
\beq
S_\be=\be^2\partial_\be F(\be)=\be^2\partial_\be F_\be
\:.\label{e}\eeq

\beq
U_\be=\be\partial_\be F(\be)+ F_\be=-e\partial_\be \ln Z_\be
\:.\label{ie}\eeq

In a similar way, one can consider  spinor fields. It is sufficient to
make use of the following formal identity \cite{dowk89-327-267}
\beq
F_f(\be)\equiv 2F_{2\be}-F_\be\:,
\label{FFE}\eeq
where on the r.h.s. the spinor quantities are left understood in the
formal expression of the bosonic free energy. As a consequence, the
horizon divergences of the bosonic sector cannot be compensated by the
corresponding fermionic ones.

\s{Spectral function for rank 1 Riemannian spaces conformally related to
$\M^N=\R^+\times\M^{N-1}$ }
\label{S:SF}

In many interesting physical cases, the Euclidean
optical metric my be written in the form (see Eq. (\ref{rl}) and Sec. 6)
\beq
d\bar{s}^2=d\tau^2+ \frac{r_H^2}{\rho^2}
\at d\rho^2+d\si^2_{N-1} \ct
\:,\label{rlo}\eeq
where $d\tau=b dx^0$, $r_H$ being a characteristic length (for example the
horizon radius),
$(r_H/\rho)^2$ the conformal factor and $d\si^2_{N-1}$ the metric of a
$N-1$-dimensional manifold .

Here we derive the spectral measure of the operator $\bar L_N$,
as defined by Eq.~(\ref{aconf}), acting on scalars in the spatial
section of the manifold defined by the metric Eq.~(\ref{rlo}).
Using such equations
(for convenience now we put $r_H=1$; in this way
all quantities are dimensionless; the dimensions will be easily
restored at the end of calculations) we easily obtain
\beq
dV&=&\rho^{-N}\,d\rho\,dV_{N-1}\:,\nn\\
\bar L_N&=&-\lap_N^{\bar g}-\ro_N^2+C\rho^2\:,\label{da}\\
\lap_N^{\bar g}&=&\rho^2\partial_\rho^2-(N-2)\rho\,\partial_\rho
+\rho^2\lap_{N-1}\nn\:,\eeq
where $\lap_{N-1}$ is the Laplace-Beltrami operator on the
manifold $\M^{N-1}$ and $dV_{N-1}$ its invariant measure.
We have also set $\ro_N=(N-1)/2$ and $C=m^2+\xi R^g$.
It should be noted the appearance of an effective "tachionic" mass
$-\ro_N^2$, which has important consequences on the structure of the
$\zeta$-function related to the operator $\bar L_N$.

In order to study the quantum properties of matter fields defined on
this ultrastatic manifold, it is sufficient to investigate the kernel
of the operator $e^{-t\bar{L}_N} $.
To this aim, we will search for the spectral resolution
of the elliptic operator $\bar{L}_N$.
Let $\Psi_{r\al}(x)$ be its eingenfunctions, namely
\beq
\bar{L}_N \Psi_{r\al}(x)=\la_r^2 \Psi_{r\al}(x)
=\la_r^2f_{\al}(\vec x)\phi_{r\al}(\rho)
\:,\label{LNPsi}
\eeq
where $f_{\al}(\vec x)$ are the (normalized) eingenfunctions
of the reduced operator $L_{N-1}=-\lap_{N-1}+C$ with eigenvalues
$\la_\al^2$. Note that we assume $C$ to be constant. This means that
we restrict ourselves to consider only manifolds $\M^D$ with constant
scalar curvature, or alternatively minimally coupled fields.
Moreover, to avoid null eigenvalues we suppose $C>0$, but the
results can be easily extended to the case $C=0$.
Note that the spectrum of $L_{N-1}$ could also be continuum.

The differential equation which determines the continuum spectrum
turns out to be
\beq
\aq\rho^2\,\partial_\rho^2
-(N-2)\rho\,\partial_\rho
-\rho^2\la_\al^2+\ro_N^2+\la_r^2
\cq\phi_{r\al}(\rho)=0
\:.\label{Bessel}\eeq
The only solutions of the latter equation
with the correct decay properties at
infinity are the Bessel functions of imaginary argument
$K_{ir}(\rho\la_\al)$ with $\la_r^2=r^2$
(if $C=0$ the operator $L_{N-1}$ has a zero mode and gives other
solutions to Eq.~(\ref{Bessel})).
Thus we have
\beq
\phi_{r\al}(\rho)=\rho^{\frac{N-1}{2}}
K_{ir}(\rho\la_\al)
\:.\label{10}\eeq

If we interpret in the sense of the distribution the
following innner product
\beq
(\Psi_{r\al},\Psi_{r'\al'})=
\int_0^\ii\frac{d\rho}{\rho^N}\int dV_{N-1}\,
\Psi^*_{r\al}(x)\Psi_{r'\al'}(x)
\:,\label{vbvb}
\eeq
we have the normalization condition
\beq
(\Psi_{r\al},\Psi_{r'\al'})=\de_{\al\al'}
\frac{\de(r-r')}{\mu(r)}
\:,\label{mu}
\eeq
where $\mu(r)$ is the spectral measure
associated with the continuum spectrum.
Thus, for the heat kernel of any suitable function $f(\bar L_N)$
we may write
\beq
<x|f(\bar L_N)|x'>
=\int_0^\ii dr\,\mu(r)\,f(r^2)
\sum_\al\Psi^*_{r\al}(x')\Psi_{r\al}(x)
\:.\label{hk2}
\eeq
The measure $\mu(r)$ may be determined
in the following standard way (Harish-Chandra's method \cite{helg84b}),
which makes use of the asymptotic behaviour of the Mac Donald
functions at the origin.
{}From Eq.~(\ref{Bessel}) and its complex conjugate
and making use of Eq. (\ref{10}) one arrives at
\beq
(\la_r^2-\la_{r'}^2)\at\phi_{r\al},\phi_{r'\al'}\ct
=\lim_{\rho\to 0}\rho^{-(N-2)}
\at\partial_\rho\phi^*_{r\al}\phi_{r'\al'}
-\phi^*_{r\al}\partial_\rho\phi_{r'\al'}\ct
\:.\label{v}
\eeq
By means of Eqs. (\ref{10}) and (\ref{mu}) we get
\beq
\frac{\de(r-r')}{\mu(r)}
=\lim_{u\to0}\frac{u}{r^2-r'^2}
\aq\partial_u\,K_{ir}^*(u)K_{ir'}(u)
-K_{ir}^*(u)\partial_uK_{ir'}(u)\cq
\:,\label{v1}
\eeq
where again, the limit has to be understood in the sense of
distributions. Recalling that for $u\to0$
\beq
K_{ir}(u)\sim\frac12\aq
\Ga(-ir)\at\frac u2\ct^{ir}
+\Ga(ir)\at\frac u2\ct^{-ir}\cq\:,
\hs ir\not\in\Z\eeq
and
\beq
\lim_{u\to 0}\frac{u^{\pm ix}}{x}=\mp\pi i\de(x)
\:,\label{asdf}\eeq
one finally has
\beq
\mu(r)=\frac{2}{\pi|\Ga(ir)|^2}=\frac2{\pi^2}\,r\sinh\pi r
\:,\label{v2}
\eeq
which is in agreement with the 2-dimensional Kontorovich-Lebedeev inversion
formula \cite{terr85b}.

Since our aim is to evaluate the trace of functions of $\bar L_N$
using Eq.~(\ref{hk}), in particular $\Tr\exp(-t\bar L_N)$,
it is convenient to make the sum over $\al$, introducing
the total spectral measure
\beq
\mu_{\bar{L}_N}(r,x)=\mu(r)\,\rho^{N-1}\sum_\al|f_{\al}(\vec x)
K_{ir}(\rho\la_\al)|^2
\label{PM}
\eeq
and integrate on the manifold defining
\beq
\mu_I(r)=\int_{{\cal M}^N}\mu_{\bar{L}_N}(r,x)\,dV=
\mu(r)\,\int_0^\ii
\sum_\al |K_{ir}(\rho\la_\al)|^2\frac{d\rho}{\rho}
\:,\label{PMI}\eeq
where we have used the normalization properties of $f_\al$.
In this way, for any suitable function $f(\bar L_N)$ we have
\beq
\Tr f(\bar L_N)=\int_0^\ii f(r^2)\,\mu_I(r)\,dr
\:.\label{TrfLN}\eeq

As a simple application of Eq.~(\ref{PM}) let us consider a massless
scalar field in a $D$-dimensional Rindler space-time.
In this case the optical spatial section turns out to be the
hyperbolic space $H^{N}$ and the measure $\mu_{\bar L_N}$
should not depend on $x$, since one is dealing with
a homogeneous space and it should coincide with the known Plancherel measure.
For this case $\M^{N-1}=\R^{N-1}$ and moreover $C=0$.
The reduced operator $L_{N-1}=-\lap_{N-1}$ has a continuum spectrum,
the eigenvalues being $k^2=\vec k\cdot\vec k$ and the corresponding
eigenfunctions $f_{\vec k}=(2\pi)^{-(N-1)/2}\exp(i\vec k\cdot\vec x)$.
As a consequence
\beq
\Phi_N(r)\equiv\mu_{\bar L_N}(r)&=&\mu(r)\,
\frac{\Omega_{N-2}\rho^{N-1}}{(2\pi)^{N-1}}
\int_0^\ii|K_{ir}(\rho k)|^2\,k^{N-2}\,dk
\nn\\&=&
\frac{2}{(4\pi)^{N/2}\Ga(N/2)}
\frac{|\Ga(ir+\ro_N)|^2}{|\Ga(ir)|^2}
\:,\label{Plan}
\eeq
$\Omega_N$ being the volume of the $N$-dimensional sphere.
Of course, Eq.~(\ref{Plan}) is the correct Plancherel measure
of the Laplace operator in $H^N$ \cite{camp90-196-1}.

\s{The heat kernel for massive fields}
\label{S:HK}

Here we derive a general expression for $\mu_I(r)$ by making use of
Eq.~(\ref{PMI}) and then derive the trace of the heat kernel,
which is needed for the construction of the partition function according
to Eq.~(\ref{logPF-Jacobi}).
Now we use the Mellin-Barnes
representation \cite{grad80b}
\beq
K_{ir}^2(\rho\la_\al)=\frac1{4i\sqrt\pi}\int_{\Re z>1}
\frac{\Ga(z+ir)\Ga(z-ir)\Ga(z)}{\Ga(z+1/2)}
\,\rho^{-2z}\la_\al^{-2z}\,dz
\label{ml}
\eeq
and observe that, for $\Re z>(N-1)/2$, the sum over $\al$ can be done
and gives
\beq
\sum_\al\la_\al^{-2z}=\ze(z|L_{N-1})
\:.\label{zred}\eeq
In  integrating over $\rho$, one has to pay attention
to the fact that the result is formally divergent.
For this aim we introduce a horizon cutoff parameter $\ep$ and,
when possible, we take the limit $\ep\to0$.
Then we get
\beq
\mu_I(r)=\frac{\mu(r)}{8i\sqrt\pi}
\int_{\Re z=c>(N-1)/2}
\frac{\Ga(z+ir)\Ga(z-ir)\Ga(z)\ze(z|L_{N-1})}
{z\Ga(z+1/2)\ep^{2z}}\,dz
\:.\label{muIr}\eeq
The integration over $z$ can be done since the meromorphic structure
of $\Ga$-and $\zeta$-functions are known. In fact, we have
\cite{mina49-1-242}
\beq
\Ga(z)\ze(z|L_{N-1})
=\sum_{n=0}^\ii\frac{K_n(L_{N-1})}{z-\frac{N-1-n}2}
+J_{N-1}(z)
\:,\label{S}\eeq
where $J_{N-1}$ is an analytic function.
Since the manifold $\M^{N-1}$ has no boundary,
all $K_n$ with odd $n$ are vanishing.

To make the integral we consider the rectangular contour
$\Ga\equiv\{\Re z=c,\Im z=a,\Re z=-c,\Im z=-a\}$ and observe that the
two horizontal paths $\Im z=\pm a$ give a vanishing contribution
in the limit $a\to\ii$, as well as the path $\Re z=-c$
in the limit $\ep\to0$. Also the poles in the strip $-c<\Re z<0$
give a vanishing contribution as soon as $\ep\to0$.
Then we have to take into consideration only the
poles of the integrand in Eq.~(\ref{muIr})
in the half-plane $\Re z\geq0$. Such a function
has simple poles at the points $z=0$, $z=-n\pm ir$ and
$z=(N-1-n)/2$ ($n\geq0$).
If $D$ is even, that is $N$ is odd, $z=0$ is a double pole.
It is clear that all poles with $\Re z>0$
give rise to divergences, the number of them depending on $N$,
while the poles at $z=0$ and $z=\pm ir$ give rise to finite
contributions. As a result one obtains
\beq
\mu_I(r)&=&\sum_{n=0}^{\aq\frac{N-2}2\cq}
\frac{K_{2n}(L_{N-1})\,\Phi_{N-2n}(r)}{N-1-2n}
\at\frac{4\pi}{\ep^2}\ct^{\frac{N-1-2n}2}
\nn\\&&\hs
+\frac{\ze'(0|L_{N-1})}{2\pi}
+\frac{\ze(0|L_{N-1})}{2\pi}\aq
\psi(ir)+\psi(-ir)-2\ln\frac\ep2-\pi\de(r)\cq
\:,\label{muIrFF}\eeq
where $\aq\frac{N-2}2\cq$ is the integer part of the number $\frac{N-2}2$,
$\psi(z)$ the logarithmic derivative of $\Ga$
and $\de(r)$ the usual Dirac $\de$-function.
Note that for even $N$, $\ze(0|L_{N-1})$ is vanishing and so the last
term in the latter equation disappears.

Now the trace of the heat kernel can be computed by using
Eq.~(\ref{TrfLN}) with $f(r^2)=\exp(-tr^2)$.
We write it in the form
\beq
\Tr e^{-t\bar L_N}&=&
\sum_{n=0}^{\frac{N-3}2}
\frac{K_{2n}(L_{N-1})\,K(t|-\lap_{H^{N-2n}}-\ro^2_{N-2n})}
{N-1-2n}\at\frac{4\pi}{\ep^2}\ct^{\frac{N-1-2n}2}
\nn\\&&\hs\hs\hs
+\frac{1}{4\sqrt{\pi t}}
\aq\ze'(0|L_{N-1})-2\ze(0|L_{N-1})\ln\frac\ep2\cq
\nn\\&&\hs\hs
-\frac{\ze(0|L_{N-1})}4
+\frac{\ze(0|L_{N-1})}{2\pi}
\int_{-\ii}^{\ii}\psi(ir)e^{-tr^2}\,dr
\:,\label{Ktodd}\eeq
\beq
\Tr e^{-t\bar L_N}&=&\sum_{n=0}^{\frac{N-2}2}
\frac{K_{2n}(L_{N-1})\,K(t|-\lap_{H^{N-2n}}-\ro^2_{N-2n})}
{N-1-2n}\at\frac{4\pi}{\ep^2}\ct^{\frac{N-1-2n}2}
\nn\\&&\hs\hs\hs\hs
+\frac{1}{4\sqrt{\pi t}}\ze'(0|L_{N-1})
\:,\label{Kteven}\eeq
valid for odd and even $N$ respectively.
Here by $K(t|-\lap_{H^{N-2n}}-\ro^2_{N-2n})$ we indicate the
diagonal heat kernel of a Laplace-like operator on $H^{N-2n}$.
Of course, it does not depend on the coordinates since
hyperbolic manifolds are homogeneous. Such a kernel is known in any
dimension \cite{byts94u-325} (see the Appendix).

As in the previous section, as a simple application of Eq.~(\ref{muIrFF}),
we again consider a massless scalar field in the $D$-dimensional Rindler
space-time. We have
$K_0(L_{N-1})=(4\pi)^{-\frac{N-1}2}V_{N-1}$, $K_n=0$ for $n>0$ and
$\ze(z|L_{N-1})=0$ for $z<(N-1)/2$.
Here $V_{N-1}$ is the volume of the manifold $\M^{N-1}$
(infinite transverse area). Then, using Eq. (\ref{Plan}), we immediately obtain
\beq
\mu_I(r)=\Phi_N(r)\,V_\ep\:,\hs
V_\ep=\frac{V_{N-1}\ep^{-(N-1)}}{N-1}
\:,\eeq
which is the integral version of Eq.~(\ref{Plan}).
Here $V_\ep$ may be considered as the volume of $H^N$.

\s{The thermodynamical quantities}
\label{S:FE}

Now it is quite straightforward to obtain the partition function and then
all the others thermodynamical quantities by means of
Eqs.~(\ref{logPF-Jacobi}) and (\ref{FE}).
Since the vacuum energy has been extensively studied in many papers,
here we concentrate our attention on the temperature dependent part of
the free energy (statistical sum) $F_\be=\bar F_\be=-\ln\bar Z_\be /\be$.
Using Eqs. (\ref{Ktodd}) and (\ref{Kteven}) we get
\beq
F_\be^{even\:D}&=&
\sum_{n=0}^{\frac{N-3}2}
\frac{K_{2n}(L_{N-1})\,{\cal F}_{N-2n}^\be}
{N-1-2n}\at\frac{4\pi}{\ep^2}\ct^{\frac{N-1-2n}2}
\nn\\&&\hs
-\frac{\pi}{12\be^2}
\aq\ze'(0|L_{N-1})-2\ze(0|L_{N-1})\ln\frac\ep2\cq
-\frac{\ze(0|L_{N-1})}4\:\frac{\ln\be}\be
\nn\\&&\hs\hs
+\frac{\ze(0|L_{N-1})}{2\pi\be}
\int_0^{\ii}[\psi(ir)+\psi(-ir)][1-e^{-\be r}]\,dr
\:,\label{FEeven}\eeq
\beq
F_\be^{odd\:D}&=&
\sum_{n=0}^{\frac{N-2}2}
\frac{K_{2n}(L_{N-1})\,{\cal F}_{N-2n}^\be}
{N-1-2n}\at\frac{4\pi}{\ep^2}\ct^{\frac{N-1-2n}2}
-\frac{\pi}{12\be^2}\ze'(0|L_{N-1})
\:,\label{FEodd}\eeq
where ${\cal F}_{N-2n}^\be$ indicates the free energy density for a
scalar field with (negative) square mass $-\ro_{N-2n}^2$
on an ultrastatic manifold with hyperbolic $H^{N-2n}$ spatial section,
which has been studied in detail in Ref.~\cite{byts94u-325}
and is given in the Appendix.

Some remarks on Eqs.~(\ref{FEeven}) and (\ref{FEodd}) are in order.
First of all, it has to be noted that the parameter
$\be$ is the inverse of the physical temperature
only if $b=1$. More generally, before to interpret
$\be^{-1}$ as the temperature in Eqs.~(\ref{FEeven}) and (\ref{FEodd}),
one has to make the substitution $\be\to b\be$. The reason is due to
the fact that, in order to write the metric (\ref{rl}) in the form (\ref{rlo}),
we have changed the time coordinate according to $\tau=b x_0$.

Independently on the manifold $\M^{N-1}$, we see that the (non renormalized)
free energy has a leading  divergence of the kind $\ep^{-(D-2)}$
proportional to the transverse area $V_{D-2}$,
since $K_0$ and ${\cal F}_N^\be$ are always non vanishing.
More generally, one has $\aq\frac{D-1}2\cq$ divergences of the kind
$\ep^{-(D-2-2n)}$ (depending on the manifold and the operator
$L_{N-1}$) and, for even $D$, also a possible logarithmic
divergence. All these divergences are also present in the expressions
of internal energy and entropy and their expressions can be obtained by
means of Eqs. (\ref{ie}) and  (\ref{e}). For example the internal
energy reads

\beq
U_\be^{even\:D}&=&
\sum_{n=0}^{\frac{N-3}2}
\frac{K_{2n}(L_{N-1})\,{\cal U}_{N-2n}^\be}
{N-1-2n}\at\frac{4\pi}{\ep^2}\ct^{\frac{N-1-2n}2}
\nn\\&&\hs
+\frac{\pi}{12\be^2}
\aq\ze'(0|L_{N-1})-2\ze(0|L_{N-1})\ln\frac\ep2\cq
+\frac{\ze(0|L_{N-1})}{4\be}
\nn\\&&\hs
-\frac{\ze(0|L_{N-1})}{2\pi}
\int_0^{\ii} r[\psi(ir)+\psi(-ir)]e^{-\be r}\,dr+U_0(\ep)
\:,\label{Eeven}\eeq
\beq
U_\be^{odd\:D}&=&
\sum_{n=0}^{\frac{N-2}2}
\frac{K_{2n}(L_{N-1})\,{\cal U}_{N-2n}^\be}
{N-1-2n}\at\frac{4\pi}{\ep^2}\ct^{\frac{N-1-2n}2}
+\frac{\pi}{12\be^2}\ze'(0|L_{N-1})+U_0(\ep)
\:,\label{Eodd}\eeq
where ${\cal U}_{N-2n}^\be$ indicates the free energy density for a
scalar field with (negative) square mass $-\ro_{N-2n}^2$
on an ultrastatic manifold with hyperbolic $H^{N-2n}$ spatial section
and $U_0(\ep)$ is the vacuum energy.

With regard to the internal energy, we have at disposal a
renormalization procedure, which is well understood for $D=4$. In
fact, in Rindler and black hole space-times it is known that the
renormalized stress-energy tensor is finite at the horizon in the
Hartle-Hawking state \cite{scia81-30-327,brow85-31-2514},
corresponding to the temperature $\be=\be_H$. This is equivalent to
write
\beq
U(\be)^{ren}=U_\ep(\be)-U_\ep(\be_H)+\mbox{finite part}\, ,
\label{uren}
\eeq
where $U_\ep(\be)$ is the divergent part of the internal energy and it
may be read off the Eqs. (\ref{Eeven}) and  (\ref{Eodd}).
Thus, the divergences are present in the expression of the
renormalized internal energy, but only for some particular value of
$\be$, say $\be_H$ ($\be_U$).
For example, in the case of Rindler space-time, such a value is
$\be_U=2\pi a^{-1}$, the Unruh temperature (here $a$ is the acceleration),
while in the 4-dimensional black hole
background one has $\be_H=8\pi MG$, the Hawking temperature.

In the general case, we may use of the same renormalization
procedure. Note, however, that the corresponding renormalized partition
function, free energy and entropy
remain divergent also at the distinguised temperature $\be=\be_H$.

\s{Some physical applications}
\label{S:PA}

As simple physical applications of the general formulae derived in
Sec.~\ref{S:FE}, here we consider the cases in which $\M^{N-1}$
is a homogeneous manifold with constant scalar curvature $\ka$.
Of course we have the three possibilities
$\M^{N-1}\equiv\R^{N-1}$ ($\ka=0$), $\M^{N-1}\equiv S^{N-1}$
($\ka>0$) and finally $\M^{N-1}\equiv H^{N-1}$ ($\ka<0$), but here we
only consider in more detail the first two cases.
The first one corresponds to the conformal treatment of
the $D$-dimensional Rindler space-time,
while the second appears when one studies the physics of black holes
near the horizon.  For $D=4$, this case has been studied in
Ref.~\cite{cogn95-12-1927}).

\ss{Statistical mechanics for massive fields
in the Rindler $D$-dimensional space-time}

As we have already observed, after a conformal transformation, the spatial
section of the
Rindler space-time is of the kind condidered in the paper.
For this special case, the curvature of $\M^{N-1}$ is vanishing ($k=0$)
and so one easily has
\beq
K_{2n}(L_{N-1})=\frac{(-m^2)^n}{n!}
\frac{V_{N-1}}{(4\pi)^{\frac{N-1}2}}
\:,\hs
\ze(z|L_{N-1})=\frac{V_{N-1}\Ga(z-\frac{N-1}2)}
{(4\pi)^{\frac{N-1}2}\Ga(z)}\,m^{N-1-2z}
\:,\label{0ok}\eeq
where $C=m^2$ has been put since Rindler is a flat manifold.
Now, using Eqs.~(\ref{FEeven}) and (\ref{FEodd}) together with
the two equations above, we obtain
\beq
F_\be^{Rind}&=&\sum_{n=0}^{\frac{N-3}2}
\frac{V_{N-1}\,\ep^{-(N-1-2n)}}{(N-1-2n)\,n!}
\at-\frac{m^2}{4\pi}\ct^n\,{\cal F}_{N-2n}^\be
\nn\\&&\hs
-\frac{\pi}{12\be^2}
\frac{V_{N-1}}{\Ga(\frac{N+1}2)}
\aq\ga+\psi(\fr{N+1}2)-\ln\fr{m^2\ep^2}4\cq
\at-\frac{m^2}{4\pi}\ct^{\frac{N-1}2}
\nn\\&&
+\frac{V_{N-1}}{\Ga(\frac{N+1}2)}
\at-\frac{m^2}{4\pi}\ct^{\frac{N-1}2}
\aq-\frac{\ln\be}{4\be}
+\frac1{2\pi\be}
\int_0^{\ii}[\psi(ir)+\psi(-ir)][1-e^{-\be r}]\,dr
\cq\:,\label{6.2}\eeq
\beq
F_\be^{Rind}&=&\sum_{n=0}^{\frac{N-2}2}
\frac{V_{N-1}\,\ep^{-(N-1-2n)}}{(N-1-2n)\,n!}
\at-\frac{m^2}{4\pi}\ct^n\,{\cal F}_{N-2n}^\be
\nn\\&&\hs\hs
-\frac{\pi}{12\be^2}
V_{N-1}\Ga(-\fr{N-1}2)
\at-\frac{m^2}{4\pi}\ct^{\frac{N-1}2}
\:,\eeq
valid for even and odd $D$-dimension respectively.
In Eq.~(\ref{6.2}) $\ga$ is the Euler constant.
The functions
${\cal F}_{N-2n}^\be$ can be computed using Eqs.~(\ref{A3}),
(\ref{Ab3}) and (\ref{Ac3}) in the Appendix.

For example, when  $D=4$, using Eq.~(\ref{Ab3-0}), the result is
\beq
F_\be^{Rind}&=&-\frac{A\pi^2}{180\be^4\ep^2}
+\frac{Am^2(1-\ln\fr{m^2\ep^2}4)}{48\be^2}\nn\\
&&\hs+\frac{Am^2}{4\pi}
\aq\frac{\ln\be}{4\be}
-\frac1{2\pi\be}
\int_0^{\ii}[\psi(ir)+\psi(-ir)][1-e^{-\be r}]\,dr
\cq
\:,\label{erto}\eeq
where the transverse area $A=V_2$ has been introduced to compare the
latter formula with well known results (see for example
Refs. \cite{suss94-50-2700,kaba94-329-46,kaba95u-06,more95u-5}).
There is agreement in the massless case, but not in the massive case,
where we also obtain a finite contribution.
We conclude this section with some remarks on renormalization.
As we have seen above, in our formalism,
massive scalar fields in Rindler space-time can be easily treated,
because the optical spatial section turns out to be the hyperbolic
space $H^3$ and the harmonic analysis on such a manifold is well known.
The formulae are particularly simple in the massless case.
For example, in 4-dimensions, the total free energy
may be chosen in the form
\beq
F^{ren}(\be)=-\frac{A}{45(8\pi)^2\ep^2}
\aq\at\frac{\be_U}{\be}\ct^4+3\cq
\:,\eeq
where $\be_U=2\pi$ is the Unruh temperature ($a=1$).
As a consequence, the entropy turns out to be
\beq S_\be=\frac{8\pi^2 A}{45\ep^2\be^3}
\eeq
and it diverges for every finite $\be$, but is zero at zero temperature
(the Fulling-Rindler state), which is correct, since we are dealing with a
pure state. At $\be=\be_U$, corresponding to the
Minkowski vacuum, we have a divergent entropy proportional to the
area, regardless of the fact that the Minkowski vacuum is a pure
state. This is also to be expected, since an uniformly accelerated
observer cannot observe the whole Minkowski space-time.
Finally with this renormalization prescription,
the internal energy should read
\beq
U^{ren}(\be)=\frac{A}{15(8\pi)^2\ep^2}
\aq\at\frac{\be_U}{\be}\ct^4-1\cq
\eeq
and this is vanishing and a
fortiori finite at $\be=\be_U$, as it should be. Furthermore, at
$\be=\ii$, namely in the Fulling-Rindler vacuum, it is in agreement
with the result obtained in Ref.~\cite{brow86-33-2840}.

\ss{Statistical mechanics for massive fields
in a $D$-dimensional black hole background}
\label{S:BH}

Here we consider in more detail the case in which $\M^{N-1}=S^{N-1}$.
To justify this choice from a physical view point,
first of all we show that, near the horizon, a $D$-dimensional black hole
may be approximated by a manifold of this kind and so,
the thermodynamics can be derived by using the formulae of Sec.~\ref{S:FE}.

The static metric describing a $D$-dimensional Schwarzschild black hole
(we assume $D>3$) reads \cite{call88-311-673}
\beq
ds^2=-\aq1-\at\frac{r_H}r\ct^{D-3}\cq\,dx_0^2+
\aq1-\at\frac{r_H}r\ct^{D-3}\cq^{-1}\,dr^2
+r^2\,d\Omega_{D-2}
\:,\label{bh}
\eeq
where we are using polar coordinates, $r$ being the radial one and
$d\Omega_{D-2}$ the $D-2$-dimensional spherical unit metric.
The horizon radius is given by
\beq
r_H=\hat{G}_D  M^{\frac1{D-3}}\:,
\hs\hat G_D=\aq\frac{2\pi^{\frac{D-3}2}\,G_D}
{(D-2)\Ga(\frac{D-1}2)}\cq^{\frac1{D-3}}
\:,\label{rh}
\eeq
$M$ being the mass of the black hole and $G_D$
the generalized Newton constant. The associated Hawking temperature
reads  $\be_H=4\pi r_H/(D-3)$. The corresponding Bekenstein-Hawking
 entropy may be computed by making use of
\beq
\be_H=\frac{\partial S_H}{\partial M}\,.
\eeq
Thus we have
\beq
S_H=4\pi \fr{\hat{G}_D}{D-2} M^{\fr{D-2}{D-3}}
\eeq
{}From now on, we put $r_H=1$.
It may be convenient to redefine the radial Schwarzschild coordinate
$r=r(\rho)$ by means of the implicit relation
\beq
\rho^2&=&\frac4{D-3}\aq e^{r-1}
\exp\int\frac{dr}{r^{D-3}-1}\cq^\frac1{D-3}\:,\nn \\
&\sim&\frac2{D-3}(r-1)e^{\frac{(D-2)(r-1)}2}+\dots
\:,\label{ro}\eeq
and time $x_0=x'_0/b$,  $b=(D-3)/2$ in order to have
$g_{00}=\rho^2+O(\rho^4)$.
In the new set of coordinates we have
\beq
ds^2=-\frac{1-r^{3-D}(\rho)}{b^2}\,dx_0'^2+
\frac{1-r^{3-D}(\rho)}{b^2\rho^2}\,d\rho^2
+r^2(\rho)\,d\Omega_{D-2}
\:,\eeq
and finally the optical metric reads
\beq
d\bar s^2 =-dx_0'^2+\frac{1}{\rho^2}\aq
d\rho^2+G(\rho)\,d\Omega_{D-2}\cq
\:,\label{OMbh}\eeq
where we have set
\beq
G(\rho)=\frac{(b\,r\,\rho)^2}{1-r^{3-D}}
=1+O(\rho^2)
\:.\eeq
{}From the latter equation we see that, near the horizon $\rho=0$,
we can set $G(\rho)=1$ and so the optical metric assumes the form
considered in previous Sections. In this approximation the manifold
$\M^{N-1}$ becomes the unit shpere $S^{N-1}$. We have
\beq
d\bar s^2 \simeq-dx_0'^2+\frac1{\rho^2}\aq
d\rho^2+d\Omega_{D-2}\cq
\:.\label{AMbh}\eeq
Such a metric can be considered as an approximation of the one of the
black hole in Eq.~(\ref{OMbh}) in the sense that, near the horizon, the
geodesics are essentially the same for both the metrics.
The metric (\ref{AMbh}) can be related to a manifold with curvature
$R^{\bar g}=-(D-1)(D-2)+O(\rho^2)$, then, according to Eq.~(\ref{Aq}),
the relevant operator becomes
\beq
\bar L_N=-\lap_N^{\bar g}-\ro_N^2+C\rho^2 +O(\rho^4)
\:,\label{ROBH}\eeq
where now $C$ is a positive constant, which takes into account of
mass and curvature contributions to this order.
Note that since for the original manifold $R^g=0$, $\xi$ does not appear
in the formulae. This effectively happens if we approximate the metric
after the optical transformation has been done. More simply,
one can put $\xi=\xi_D$ in Eq.~(\ref{Aq}).

The discussion for arbitrary $D=N+1$ is quite involved even though it may
be done, since the $\ze$-functions of the Laplace-Beltrami operators on
$S^{N-1}$ are known (see the Appendix).

As a more explicit example, now we consider a scalar field in
a 4-dimensional Schwarzschild background.
Using these techniques, such a case has been considered in
Ref.~\cite{cogn95-12-1927}, where we refer the interested reader for more
details.
We have $r_H=2MG$, $b=1/2$,
\beq
\rho=2(r-1)^{\frac12}e^{(r-1)/2}
\:,\label{ro1}\eeq
and
\beq
r=1+\frac{\rho^2}{4}-\frac{\rho^4}{16}+O(\rho^6)
\:.\label{456}\eeq
Then, according to Eq.~(\ref{ROBH}), the relevant operator becomes
\beq
\bar L_3=-\bar\lap_3-1+C\rho^2
\:,\eeq
where $C=m^2+1/3$ takes into account of the curvature
$R^{\bar g}=-6+2\rho^2$ of the optical manifold.

Now, directly using Eqs.~(\ref{logPF-Jacobi}), (\ref{FE}),
(\ref{FEeven}) and (\ref{Ab3-0}), after
the replacement $\be\to\be/2$ due to the redefinition of the
Schwarzschild time (remember that $b=1/2$), for the total free energy
we obtain
\beq
F^{bh}(\be) &=&-Aj_\ep+\frac14\ze(-\fr12|\bar L_3)
-\frac{2\pi^2A}{45\ep^2\be^4}-\frac{A}{12\be^2}\aq
\frac{\ze'(0|L_2)}2-\ze(0|L_2)\ln\frac\ep2\cq \nn\\&&
-\frac{A\ze(0|L_2)}{16\pi\be}\ln\frac\be2
+\frac{A\ze(0|L_2)}{8\pi^2\be}\int_0^\ii
\ln\at1-e^{-\be r/2}\ct\, \aq\psi(ir)+\psi(-ir)\cq\,dr
\:,\label{bhf} \eeq
where we have written the Jacobian contribution to the partition
function due to the conformal transformation in the form $A\be j_\ep$,
and now $A=4\pi r_H^2$ is the transverse area of the black hole.
The $\zeta$-function related to the operator $L_2$ on the sphere is
given in the Appendix by Eq.~(\ref{99}) with $C=m^2+1/3$,
then $\ze(0|L_2)=m^2$.

The leading divergence, due to the optical volume,
is proportional to the horizon area \cite{thoo85-256-727},
but in contrast with the Rindler case, a
logarithmic divergence is also present, similar to the one found in
Refs.~\cite{solo94u-246,deal94u-347}.
This is a feature of even dimensions (see Sec.~\ref{S:FE}).

Let us briefly discuss the renormalization of the internal
energy in this particular case.
We recall that one needs a renormalization in order to remove the
vacuum divergences. These divergenges, as well as the
Jacobian conformal factor,  do not contribute to the entropy.
However the situation presented here is complicated
by the presence of horizon divergences, controlled by the cutoff parameter
$\ep$. In the 4-dimensional Schwarzschild space-time, it is known that the
renormalized stress-energy tensor is well defined at the horizon in
the Hartle-Hawking state \cite{scia81-30-327,brow85-31-2514}, which in
our formalism corresponds to the Hawking temperature $\be=\be_H$.
The renormalized internal energy reads (the dots stay for finite
contributions at the horizon, which we do not write down because their
value depend on the approximation made)
 \beq
U^{ren}(\be)&=&\frac{A}{30(8\pi)^2\ep^2}
\aq\at\frac{\be_H}{\be}\ct^4-1\cq
-\frac{A}{3(8\pi)^2}\ln\ep \aq\at\frac{\be_H}{\be}\ct^2-1\cq
\:\:+\dots \:,\eeq
which has no divergences for $\be=\be_H$, the Hawking temperature,
while the entropy
\beq
S_\be=\frac{8\pi^2A}{45\ep^2\be^3}
-\frac{A\ln\ep}{6\be}\:\:+\dots\:,\label{S1}\eeq
also for $\be=\be_H$ contains the well known divergent
term proportional to the horizon
area \cite{thoo85-256-727} and, according to Ref.~\cite{solo94u-246},
a logarithmic divergence too. Eq.~(\ref{S}) is vanishing in the
Boulware vacuum corresponding to $\be=\ii$.

{}From this renormalization procedure we get for the renormalized black hole
free energy
\beq
F^{ren}(\be) &=&-\frac{A}{90(8\pi)^2\ep^2}
\aq\at\frac{\be_H}{\be}\ct^4+3\cq
+\frac{A}{3(8\pi)^2}\ln\ep
\aq\at\frac{\be_H}{\be}\ct^2+1\cq\nn\\
&&\hs -\frac{A}{12\be^2}\aq
\frac{\ze'(0|L_2)}2+m^2\ln2\cq
-\frac{Am^2}{16\pi\be}\ln\frac\be2\nn\\
&&\hs\hs +\frac{Am^2}{8\pi^2\be}
\int_0^\ii\ln\at1-e^{-\be r/2}\ct\,
\aq\psi(ir)+\psi(-ir)\cq\,dr
\:.\label{bhf2} \eeq

In the general case, the discussion is quite similar and it  can be
performed by using Eqs.~(\ref{FEeven}) or (\ref{FEodd})
with the replacement $\be\to(D-3)\be/2$.

\s{Conclusions}
\label{S:C}

In this paper the first quantum corrections to the thermodynamic
quantities of fields in a
$D$-dimensional Rindler-like space have been investigated
making use of conformal transformation techniques and $\zeta$-function
regularization.
In this way, we have worked within the so
called optical manifold, which is ultrastatic, and the
use of finite temperature methods is quite straightforward.

The general form of the
horizon divercences of the free energy has been obtained as a function of
free energy densities of fields having negative square masses (absence
of the gap in the Laplace operator spectrum) on ultrastatic manifolds
with hyperbolic spatial section $H^{N-2n}$ and of the Seeley-DeWitt
coefficients $K_{2n}(L_{N-1})$ of the Laplace operator on $\M^{N-1}$.
Since there exists recurrence relations for free energy densities
(see the Appendix), it is sufficient to study the cases $D=3$ and $D=4$
($D=4$ and $D=5$ for applications to black holes).
The leading divergence can be seen to be given by
 the volume of the spatial section of the optical manifold.
A finite contribution is
also obtained and this depends on $\ze(0|L_{N-1})$ and on its first
derivative. For $D=4$, our results are consistent with the ones obtained
with brick wall, path-integral and canonical methods
\cite{ghos94-73-2521,deal95u-33,barb95u-155}.

With regard to physical applications, we have used the general results
on finite temperature field theory
in order to investigate the quantum corrections to the
Bekenstein-Hawking entropy for massive fields in a
large mass black hole background.
This approach gives rise to a leading divergence for the entropy
similar to the one obtained for the Rindler case background,
but in this case other divergent contributions are present
and their structure depend on the dimension of the
space-time considered.
Here we have shown how it is possible to get the general form valid for
an arbitrary dimension and we have explicitly considered the case
$D=4$.

We also would like to mention the results obtained in
Ref.~\cite{frol93-48-4545}, where the contributions to the $4$-dimensional
black hole entropy due to modes located inside and near the horizon have been
evaluated using a new invariant statistical mechanical definition for
the black hole entropy. The finite contributions, namely the ones
indipendent on the horizon cutoff, are compatible with our results.

As far as the horizon divergences are concerned we recall that they may
be interpreted physically in terms of the infinite gravitational redshift
existing between the spatial infinity, where one measures the generic
equilibrium temperature and the horizon, which is classically
unaccessible for the Schwarzschild external observer.
Furthermore, we have argued that they are absent in the
internal energy at the Unrhu-Hawking temperature. However, they remain in
the entropy and in the other thermodynamical quantities, as soon as
one assumes the validity of the usual thermodynamical relations. For
$D=4$, a possible way to deal with such divergences has been suggested in
Refs.~\cite{thoo85-256-727,frol93-48-4545,frol94-74-3319}, where it has
been argued that the quantum fluctuations at the horizon might provide
a natural cutoff. In particular, choosing the horizon cutoff parameter
of the order of the Planck length ($\ep^2\sim G$), the leading "divergence",
evaluated at the Hawking temperature, turns out to be of the form of
the the "classical" Bekenstein-Hawking entropy. This seems a
reasonable assumption, because we have worked within the fixed
background approximation. However one should
remark that other terms are present, giving contributions which
violate the area law. A more elaborate discussion for $D=4$ can be
found in Ref.~\cite{frol94u-211}.
Alternatively, one may try to relate the horizon
divergences to the ultraviolet divergences of quantum gravity, thus
arriving at the theory of superstring progagating in a curved
space-time \cite{suss94-50-2700} or at the renormalization group approach
\cite{odin95u-27}.

Finally, we  mention that there has been  the proposal to absorbe
the horizon divergences, at least for $D=2,3,4$, by making use of
the standard ultraviolet gravitational constant renormalization
\cite{suss94-50-2700,solo95-51-609,deme95u-3,furs94u-20,lars95u-66}.
This proposal is essentially based on the use of Euclidean section
with a conical singularity and
associated heat kernel expansion. The problematic issue consisting in
dealing with a finite temperature theory in a non ultrastatic space-time
is solved working within a non vanishing
conical singularity and interpreting the deficit angle of the
Euclidean compactified time  as the inverse of the
equilibrium temperature (the absence of the conical singularity gives
the correct Hawking temperature). However, the resulting partition function
has, apparently, a wrong dependence on this "temperature".
Furthermore, it seems that the
only divergences present are the usual ultraviolet ones
associated with the definition of the partition function as determinant
of an elliptic operator. These divergences are then absorbed in the
gravitational constant renormalization. However, the naive use of
$\zeta$-function regularization should get rid off these ultraviolet
divergences.
Thus it seems to exist a disagreement between this approach and our
approach based on the conformal transformation techniques. It is our
opinion that this disagreement might depend on a non commutative
property present in the evaluation of heat kernel trace on a cone. It
should be interesting to elucidate this issue.

\ack{We wish to thank E.S. Moreira Jnr. for pointing out a mistake
in Eq. (\ref{erto}) in the first version of the manuscript
and L. Vanzo for discussions.}

\appendix

\app{Heat kernel, $\zeta$-function and free energy in constant
curvature spaces}

To start with, let us write down a representatation for the
statistical sum which may be useful for investigating the high temperature
expansion. For  detailed derivations we refer the reader to
Ref.~\cite{byts94u-325}, where all material of this Appendix can be
found. It reads
\beq
{\cal F}_{N}^\beta=-\frac{1}{\pi i}\int_{\Re z=c}dz \ze(z)_R \Ga(z-1)
\ze(\frac{z-1}{2}|A_N)\be^{-z}\,,
\label{M}\eeq
where $A_N=-\lap_N+\al^2+\ka\ro_N^2$,  acting on fields in $H^N$ or
$S_N$, with $\al$ a constant which may
go to zero.

\ss{Hyperbolic manifolds}
Here we only report the equations concerning heat kernel,
$\zeta$-function and free energy for scalars fields in hyperbolic
manifolds, we need in the paper.

The measure for the Laplace-Beltrami operator acting on scalar fields
in $H^N$ satisfy the recurrence relations
\beq
\Phi_{N+2}(r)=\frac{\ro_N^2+r^2}{2\pi N}\Phi_N(r)
\:,\eeq
which permit to derive recurrence formulae for all others quantities
one is interested in. In particular, for the operator
$A_N$ (here $\ka=-1$) we have
\beq
K(t|A_{N+2})=-\frac1{2\pi N}\aq\partial_t+\al^2+\ka\ro_N^2\cq K(t|A_N)
\:,\label{A1}\eeq
\beq
\tilde\ze(s|A_{N+2})=-\frac1{2\pi N}
\aq(\al^2+\ka\ro_N^2)\tilde\ze(s|A_N)-\tilde\ze(s-1|A_N)\cq
\:,\label{A2}\eeq
where $K(t|A_N)$ and $\tilde\ze(s|A_N)$ are densities.
Finally, for the free energy density, one gets
\beq
{\cal F}_{N+2}^\beta
=-\frac1{2\pi N}\left[(\al^2+\ka\varrho_N^2)
{\cal F}_N^\beta-\tilde\zeta'(-1|A_N)\right]
\:.\label{A3}\eeq
Using equations above, one obtains the desired quantities
for fields in $H^N$,
starting from $H^3$ or $H^2$ according to whether $N$ is odd or even.
For the space $H^3$ we have
\beq
K(t|A_3)=\frac{e^{-t\al^2}}{(4\pi t)^{3/2}}
\:,\label{Ab1}\eeq
\beq
\tilde\zeta(s|A_3)=\frac{\Gamma(s-\frac32)}
{(4\pi)^{\frac32}\Gamma(s)}\,\alpha^{3-2s}
\:,\label{Ab2}\eeq
\beq
{\cal F}_3^\beta=
-\frac1{\pi i}\int_{\Re z=c}dz \ze_R(z)\Ga(z-1)
\frac{\Ga(\frac z2-2)}{\Ga(\frac{z-1}2)}\be^{-z}
\frac{\al^{4-z}}{(4\pi)^{3/2}}
\label{Ab3}\,,
\eeq
or equivalently
\beq
{\cal F}_3^\beta=-\frac{\alpha^4}{2\pi^2}\sum_{n=1}^\infty
\frac{K_2(n\beta\alpha)}{(n\beta\alpha)^2}
\:,\label{Ab31}\eeq
while for $H^2$
\beq
K(t|A_2)=\frac{e^{-t\alpha^2}}{4\pi t}
\int_{0}^{\infty}\frac{\pi e^{-tr^2}}{\cosh^2\pi r}\;dr
=\frac{e^{-t\alpha^2}}{4\pi t}
\sum_{n=0}^\infty\frac{B_{2n}}{n!}\left(2^{1-2n}-1\right)\,t^n
\:,\label{Ac1}\eeq
\beq
\tilde\zeta(s|A_2)=\frac{\alpha^{2-2s}}{4\pi(s-1)}
-\frac1\pi\int_{0}^{\infty}
\frac{r\,(r^2+\alpha^2)^{-s}}{1+e^{2\pi r}}\,dr
\:,\label{Ac2}\eeq
\beq
{\cal F}_2^\beta=-\frac1{\sqrt{2\pi}}
\sum_{n=1}^\infty\int_{0}^\infty
\frac{K_{3/2}(n\beta\sqrt{r^2+\alpha^2})\,(r^2+\alpha^2)^{3/4}}
{(n\beta)^{3/2}\,\cosh^2\pi r}\,dr
\:.\label{Ac3}\eeq
In Eq.~(\ref{Ac1}), $B_n$ are the
Bernoulli numbers and the series is convergent for $0<t<2\pi$.

Note that in the paper the relevant operator concerning
hyperbolic metric is $A_{N-2n}$ with $\al=0$.
In this limit, equations above notably simplify, especially
in the odd $N$ cases. For example, from Eq.~(\ref{Ab3})in the limit
$\al\to0$ we have
\beq
{\cal F}_3^\beta=-\frac{\ze_R(4)}{\pi^2\be^4}=-\frac{\pi^2}{90\be^4}
\:.\label{Ab3-0}\eeq

\ss{Spheres}
As has been shown in Ref.~\cite{byts94u-325},
the recurrence relations (\ref{A1}-\ref{A3}) are also valid for manifolds
with positive constant curvature ($\ka=1$). This means that for spheres,
heat-kernel, $\zeta$-function and free energy can be obtained by the
knowledge of the same quantities for $S^1$ and $S^2$.
In the paper we only need heat-kernel coefficients and $\zeta$-function.
For the circle $S^1$ we have the representation
\beq
K(t|A_1)=\frac{e^{-t\alpha^2}}{\sqrt{4\pi t}}
\sum_{n=-\ii}^{\ii}e^{-\pi^2n^2/t}
\:,\eeq
\beq
\tilde\zeta(s|A_1)=\frac{\Ga(s-\frac12)\alpha^{1-2s}}
{\sqrt{4\pi}\Ga(s)}
+\frac{2\alpha^{1-2s}\sin\pi s}\pi
\int_{1}^{\infty}
\frac{(r^2-1)^{-s}}{e^{2\pi\al r}-1}\,dr
\:,\eeq
while for $S^2$
\beq
K(t|A_2)&=&\frac{e^{-t\al^2}}{8i t}
\int_{\Ga}\frac{e^{-tz^2}}{\cos^2\pi z}\,dz
\:,\eeq
\beq
\tilde\zeta(s|A_2)=\frac1{8i(s-1)}
\int_{\Ga}\frac{(z^2+\al^2)^{-(s-1)}}{\cos^2\pi z}\,dz
\:,\eeq
where $\Ga$ is an open path in the complex plane going (clockwise)
from $\ii$ to $\ii$ around the positive real axis enclosing the point
$z=\ro_2=1/2$ (see Ref.~\cite{byts94u-325}).
Note that in the paper the operator concerning spheres is $L_{N-1}$,
so we have to choose $\al^2=C-\ro_{N-1}^2$.
Of course, other representations are at disposal and can be found for
example in Refs.~\cite{acto87-20-927,camp90-196-1,byts94u-325}.
For example,
\beq
\ze(s|L_2)=2\sum_{n=0}^\ii\frac{(-1)^n\Ga(s+n)\at C-\frac1{4}\ct^n}
{\Ga(n+1)\Ga(s)}\,\ze_H(2s+2n-1,1/2)
\:,\label{99}
\eeq
whence $\ze(0|L_2)=C-1/3$.

\end{document}